\documentclass[pra,aps,twocolumn,superscriptaddress,showpacs]{revtex4}
\usepackage{graphicx}
\usepackage{amsmath}

\newcommand{\eq}{\begin{equation}}
\newcommand{\fine}{\end{equation}}

\begin{document}

\title{

\bf \LARGE  Arbitrarily large violations of non-contextuality in single mode photon states with positive Wigner function}
\vspace{.2cm}
\date{27 March, 2018}

\author{S. M. Roy}
\email{smroy@hbcse.tifr.res.in} \affiliation{HBCSE,Tata Institute of Fundamental Research, Mumbai}

\begin{abstract}
Banaszek, W\'odkiewicz and others (\cite{Banaszek},\cite{Chen},\cite{Chen-Zhang})  made the surprising discovery that  
Einstein-Bell locality inequalities  can be  violated  
  by the two mode squeezed vacuum by  a factor $\sqrt{2}$, in spite of the fact that the state has a positive Wigner function. 
   I use here the more general Gleason-Kochen-Specker assumption of non-contextuality \cite{Gleason} to express classicality.
    I then derive non-contextuality Bell inequalities 
 for correlations of $N$ pseudo spins embedded in an infinite dimensional continuous variable Hilbert space , and show that their maximum 
 possible quantum violation is by a factor $2^{(N-1)/2}$. I find quantum states for which this maximum violation is reached.
  I also show that the familiar displaced squeezed vacuum for a single optical mode, which has 
a positive Wigner function, can violate the inequality by a factor $0.842 (\sqrt{2} )^{N-1} $ for odd $N \geq 3$ .
The arbitrarily large non-classicality  means that realizations of the pseudo-spin measurements even in a single mode photon 
state might afford similar opportunities in quantum information tasks as entangled $N$ qubit systems with large $N$.

\end{abstract}
 
\pacs{03.65.-W , 03.65.Ta ,04.80.Nn}

\maketitle

{\bf Introduction}.
Bell \cite{Bell1966} pointed out that  Gleason's theorem \cite{Gleason} of impossibility of dispersion free quantum states 
is based on a fundamental `non-contextuality'hypothesis :`that measurement of an 
observable must yield the same value independent of what other measurements may be made simultaneously'.When states are not dispersion free, 
this assumption can be changed into the less restrictive `stochastic non-contextuality hypothesis': that measured probability distribution of 
an observable must be independent of what other measurements may be made simultaneously. When the commuting observables are space-like separated , 
this is the same as Einstein's local reality hypothesis \cite{Einstein1933} and leads to Bell's theorem \cite{Bell1964} 
that quantum mechanics violates Einstein locality. There is now a definite realisation that contextuality (including non-locality 
induced by entanglement) is a valuable resource \cite{resource} for quantum computation.

Bell correlations may be generally defined as linear combinations of correlations between jointly measurable (and hence 
commuting ) observables, different terms of the linear combination being mutually non-commuting.Even when the commuting observables involved 
in Bell correlations are not spacelike separated, non-contextuality implies inequalities on these correlations. 
{\bf Quantum violation of non-contextuality is a decisive signal of non- classical behaviour}.

Einstein locality violations concern two or more spacelike separated systems, the maximum violation being by a factor upto $ \sqrt{2}$   
for two particles of spin $1/2$ \cite{Bell1964} or general spin $S$ \cite{Gisin} , and by a factor upto $2^{(N-1)/2}$ for $N$ qubits 
(\cite{mermin1990},\cite{roy-singh1991}.\cite{ardehalli1992}). In contrast,violations of non-contextuality can occur  even for a single system 
with Hilbert space of dimension $\geq 3$ \cite{Gleason}; e.g. for a particle of spin $S$ with $2S+1=2^n$, and $n$ odd, non-contextuality 
violations by a factor of $\sqrt{(2S+1)/2 }$ has been demonstrated \cite{Roy-Singh1993}.

{\bf Phas space Bell inequalities}.There has been considerable theoretical and experimental progress on quantum optical continuous variable 
EPR systems \cite{braunstein},\cite{Braunstein-Loock},\cite{Ou}.Many years ago we \cite{AMRS} derived  
phase space Bell inequalities for a four dimensional phase space .E.g. assuming existence of a positive phase space 
density $\rho(\vec{x},\vec{p})$ reproducing the four  experimental probability densities for $(q_1,q_2),(q_1,p_2),(p_1,q_2),(p_1,p_2) $ 
as marginals,
\begin{eqnarray}\label{AMRS}
&& |\int  d\vec{x}d\vec{p}\>\rho(\vec{x},\vec{p})\big[sqn F_1(q_1) sqn F_2(q_2)\nonumber\\
&& + sqn F_1(q_1) sqn G_2(p_2) +sqn G_1(p_1) sqn F_2(q_2)\nonumber\\
&& -sqn G_1(p_1) sqn G_2(p_2)\big]| \leq 2 ,
 \end{eqnarray}
 where, $F_1,F_2,G_1,G_2$ are arbitrary non-vanishing functions, which {\bf need not be periodic}. The corresponding quantum 
 inequalities , with the phase space variables replaced by operators , and phase space averages replaced by 
 quantum expectation values , are necessarily obeyed by quantum states with positive Wigner functions; but there exist states for which 
they are  violated . Optimisation of these inequalities for experiments have been considered in \cite{Grangier}.

Following a surge of interest in quantum information applications of modular observables ,
i.e. periodic functions of  phase space variables( \cite{modular},\cite{Cabello},\cite{Walborn},\cite{Asadian},\cite{Ketterer}),
Arora and Asadian \cite{Asadian} obtained for a state with a positive Wigner function,
\begin{equation}
 |Tr \rho A_1(A_2+A_2')  |+|Tr \rho A_1'(A_2-A_2')  | \> \leq 2\>,
\end{equation} 
where $A_1,A_1',A_2,A_2'$  are observables whose Wigner transforms $ A_1(q_1,p_1), A_1'(q_1,p_1), A_2(q_2,p_2),
 A_2'(q_2,p_2)$ are of magnitude  $\leq 1$. They discussed practical measurements on states violating this inequality using 
 modular observables. Comparison with the similar inequality (\ref{AMRS}) suggests that their investigations 
 may be extended to non-modular observables also.

{\bf EPR wave function}.The above results correspond to Bell's remarks on measurements of linear combinations 
of position and momentum using the EPR wave function. He concluded that the original (non-normalizable) EPR wave function leads 
to a (non-normalizable) positive Wigner function and therefore has no non-locality problem \cite{Bell1986}. 
A significant achievement of  Banaszek, W\'odkiewicz and others (\cite{Banaszek},\cite{Chen}) 
was to show that this was incorrect . They demonstrated the non-locality of a normalizable 
EPR-like state, the two-mode squeezed vacuum or NOPA (non-degenerate optical parametric amplifier) state which has a positive Wigner function.
They showed that locality inequalities on Bell correlations of phase-space displaced parity operators or of pseudo-spin observables 
for this sysystem are violated by the quantum correlations 
by a factor $ \sqrt{2}$. Thus, there exist observables for which locality need not be connected to the positivity of the Wigner function.

{\bf Present work}. Here, I shall demonstrate  that even for  a continuous variable system with only one degree of freedom, and positive Wigner function, 
non-contextuality inequalities on Bell correlations may be violated by an arbitrarily large factor. The state may be as simple as a  
squeezed coherent state. Relevant correlation measurements and quantum information applications may be possible using simple techniques of 
continuous variable quantum computation such as balanced homodyne measurements and unitary phase space displacement operations 
on electromagnetic quadratures. In fact, it has been claimed that ``this simplicity and the high efficiency when measuring and manipulating 
the continuous quadratures are the main reason why continuous-variable schemes appear more attractive than those based on discrete variables 
such as the photon number.''( Braunstein and van Loock in \cite{Braunstein-Loock}).

My first step will be to define $N$ qubit pseudo-spin operators in a single continuous variable Hilbert space, viz. quantum optical 
quadratures for a single mode.

{\bf N qubit pseudo-spin operators}. A single electromagnetic mode  of frequency $\omega$ corresponds to 
an oscillator Hamiltonian with ground state energy subtracted,
\begin{equation}
 H=\hbar \omega \hat{a}^\dagger \hat{a},
\end{equation}
where the annihilation operator $\hat{a}$ may be expressed in terms of dimensionless hermitian quadrature operators $\hat{x},\hat{p} $,
\begin{eqnarray}
  \hat{x}= (\hat{a} +\hat{a}^\dagger )/\sqrt{2};\>\hat{p}= (\hat{a} -\hat{a}^\dagger )/(i\sqrt{2})\nonumber\\
  \hat{a} =( \hat{x}+i\hat{p})/\sqrt{2} \>;\>[\hat{a},\hat{a}^\dagger]=1,\>[\hat{x},\hat{p}]=i\>.
\end{eqnarray}
I divide the configuration space $x\in R=(-\infty,+\infty)$ of eigenvalues of the quadrature $\hat{x}$ into discrete intervals 
$(aL(s-\frac{1}{2}) ,aL(s+\frac{1}{2}) ),$ 
each of length $La$ and centred at $aLs$ where $s$ is an integer,  zero, positive or negative; and 
$L=2^N$. The arbitrary parameter `a' is not to be confused with the operator $\hat{a} $.
With a view to corresponding to $2^N$ basis states of $N$ qubits, I sub-divide the $s$-th interval 
into $2^N$ sub-intervals labelled by $m$, each  of length $a$ ,
\begin{eqnarray}
 I_{m,s}= (aL(s-\frac{1}{2}) +ma, aL(s-\frac{1}{2})  +(m+1)a ),\nonumber\\
 s=integer , \>L=2^N \>,m=0,1,..,L-1 \>.
\end{eqnarray}
Then,
\begin{eqnarray}
&&\int_{ I_{m,s }} dx |x><x|=\int_{y=0}^{a} dy |m,s,y><m,s,y|\>,\\
&&\nonumber\\
&& |m,s,y>\equiv |aL(s-1/2)+am+y> .\label{xdefinition}
\end{eqnarray}
The  completeness of the states $|x>$  and orthonormality relations become 
\begin{eqnarray}\label{completeness}
&& {\bf 1} =\sum_{s=-\infty}^\infty \sum_{m=0,1,..}^{(2^N-1 ) }\int_{y=0}^{a} dy |m,s,y><m,s,y|,\\
&& \nonumber\\
&&\nonumber\\
&& <m,s,y|m',s',y'>=\delta_{s,s'} \delta_{m,m'} \delta (y-y')\>.\label{orthonormality}
\end{eqnarray}

I want to define $N$ pseudo-spin operators $\sigma_z ^{(j)}$ and the corresponding $2^N$ eigenstates 
with eigenvalues $m_j=\pm 1$,for $ j=1,2,..N$. Like  the set  $(m_1,m_2,..,m_N) $, the integer $m$ takes $2^N$ values.
 If I define,
\begin{equation} \label{m}
 m(m_1,m_2,..,m_N )=\sum_{j=1}^N 2^{(j-1 ) } (1+m_j )/ 2,
\end{equation}
then the resulting values of $m$ are $0,1,..2^N-1$, with $m=0$ for all $m_j=-1$ and $m=2^N-1$ for all $m_j=+1$. For each $m$,
the relation can be inverted to solve uniquely for $ (m_1,m_2,..,m_N )$, i.e. for $ m_j=\pm 1  $
the correspondence 
\begin{equation}\label{correspondence}
 m \>\leftrightarrow \>  (m_1,m_2,..,m_N) 
 \end{equation}
 is one-to-one. This is obvious from Eq. (\ref{m} )because $$ ((m_N +1)/2)((m_{N-1} +1)/2)...((m_1 +1)/2)$$
is just the binary representation of $m$, each of the $N$ digits being $0$ or $1$.
\\
\begin{figure*}[!]
\begin{center}
 \includegraphics[width=1.9 \columnwidth]{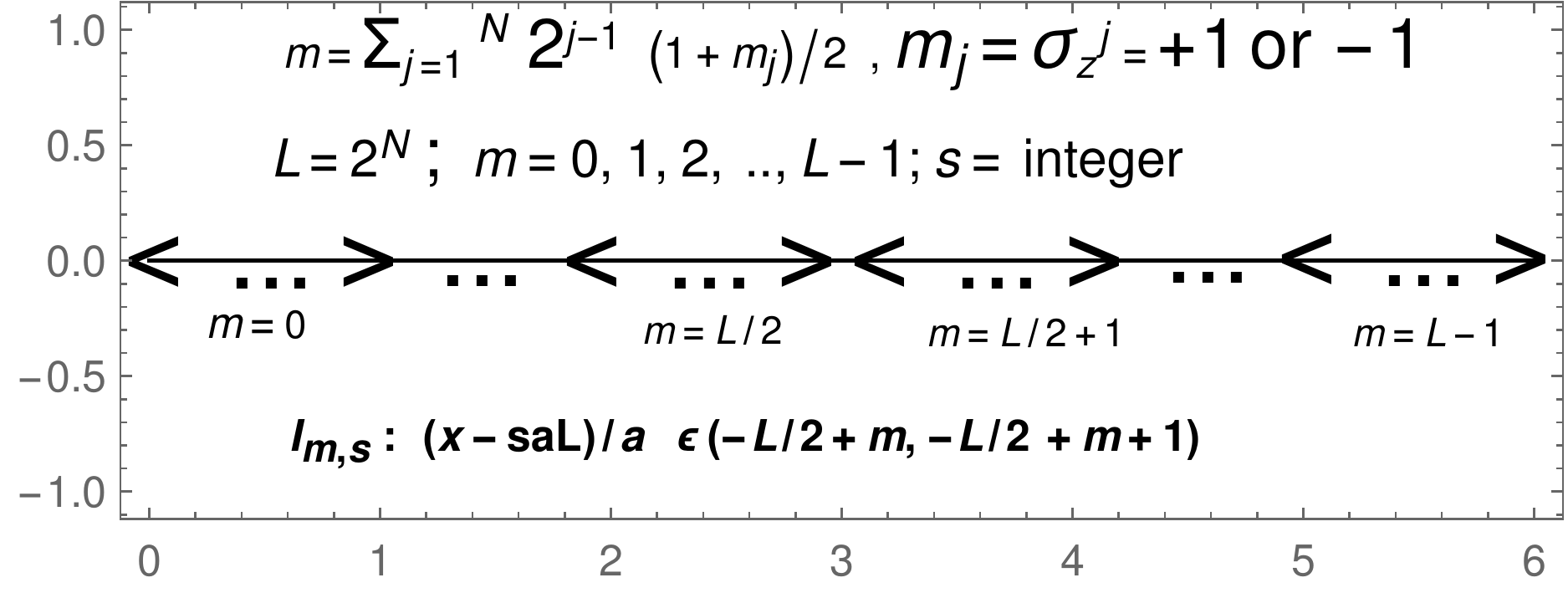}
 \caption {This figure illustrates the correspondence of the $N-$ qubit variables $m(m_1,m_2,..,m_N)$
 with configuration space intervals $I_{m,s}$, where $s$ is an integer, negative, zero or positive.
 Because of the one to one correspondence $m \>\leftrightarrow \>  (m_1,m_2,..,m_N)$,  each interval 
 also corresponds to a  definite vaue of the set $(m_1,m_2,..,m_N)$.The state (\ref{m=0,L-1}) showing 
 maximal violation of non-contextuality is a superposition of $m=0$ and $m=L-1$ states; the $m=L/2,m=L/2-1$ 
 intervals make important contributions to the single mode squeezed vacuum state (\ref{SMSV} ). } 
\end{center}
\end{figure*}
\\
Due to the one-to-one correspondence (\ref{correspondence}), we may write the orthonormality relation (\ref{orthonormality})for the states as
\begin{eqnarray}
 |m,s,y>\equiv |m_1,m_2,..,m_N;s,y>\>,\nonumber\\
 |m',s,y>\equiv |m'_1,m'_2,..,m'_N;s,y>\>, \nonumber\\
 <m,s,y|m',s',y'>=\delta_{s,s'} \delta (y-y') \prod_{j=1}^N\delta_{m_j,m'_j} .
\end{eqnarray}

For a given $m$, I now define the sub-intervals $I_{m,s}$ for all $s$ to correspond to an eigenstate of $\sigma_z^{(1)},...\sigma_z^{(N)}$,
i.e. 
\begin{eqnarray}
&&\sigma_z^{(j)} |m,s,y>=m_j |m,s,y>,  \>for \>all\> j\>and\>any\>s,y,\nonumber\\
&& M\equiv \sum_{j=1}^N 2^{(j-1 ) } (1+\sigma_z^{(j)} )/ 2\>,\nonumber\\
&& M|m,s,y>=m|m,s,y>\>.
\end{eqnarray}
Eqn. (\ref{xdefinition}) shows that the operator $M$ is  periodic in $\hat{x}$ with period $a 2^{N}$.
 Next ,the raising and lowering operators  $\sigma_\pm ^{(j)}$ should convert a state with $m_j=\mp 1$ into a state with 
 $m_j=\pm 1$, leaving all other  $m_{j'}, j'\neq j$ unchanged, and should annihilate states with $m_j=\pm 1$. The definition of $m$, Eq.(\ref{m}), 
 shows that if all $m_{j'}, j'\neq j$ are unchanged, the value of $m$ for $m_j=1$ is greater than it's value for $m_j=-1$ by $ 2^{(j-1)}$.Hence,
 \begin{equation}
 |m_1,..,m_{j-1},-m_j,..,m_N;s,y>= |m - 2^{(j-1)}m_j,s,y> \>,
\end{equation}
 and I stipulate that,
\begin{eqnarray}
& & ( \sigma_x^{(j)}\pm i \sigma_y^{(j)})|m,s,y>\nonumber\\
&=& (1\mp m_j)|m - 2^{(j-1)}m_j,s,y>\>.
\end{eqnarray}
Equivalently,
\begin{eqnarray}
 \sigma_x^{(j)}|m,s,y>&=&|m - 2^{(j-1)}m_j,s,y> ,\nonumber\\
 \sigma_y^{(j)}|m,s,y>&=& i m_j|m - 2^{(j-1)}m_j,s,y> .
\end{eqnarray} 
If $\vec{n}$ are unit vectors, these definitions may be summarized by
\begin{eqnarray}\label{sigma^j}
&& <m',s',y'|\vec{\sigma}^{(j)}.\vec{n}|m,s,y> \nonumber\\
&&=<m'_j|\vec{\sigma}.\vec{n}|m_j >\delta_{s,s'} \delta (y-y') \prod_{l\neq j}\delta_{m_l,m'_l} ,
\end{eqnarray}
where $\vec{\sigma} $ without the superscript $j$ denote the usual Pauli matrices, and $|\pm 1 >$ denote eigen vectors of $ \sigma_z $ .
The standard commutation rules follow,
\begin{equation}
 \sigma_x^{(j)} \sigma_y^{(j)}=-\sigma_y^{(j)} \sigma_x^{(j)}=i\sigma_z^{(j)} ,
\end{equation}
operators for different $j$ being mutually commuting on account of Eqn. (\ref{sigma^j} ).

Using the completeness relations (\ref{completeness}) ,these definitions are equivalent to ,

\begin{eqnarray}
\sigma_{x}^{(j)}&=&\sum_{s,m}\int_{y=0}^{a} dy |m,s,y><m - 2^{(j-1)}m_j,s,y|,\nonumber\\
\sigma_{y}^{(j)}&=&-i\sum_{s,m}m_j\int_{y=0}^{a} dy |m,s,y><m - 2^{(j-1)}m_j,s,y|,\nonumber\\
\sigma_{z}^{(j)}&=&\sum_{s,m} \int_{y=0}^{a} dy \>m_j |m,s,y><m,s,y|.
\end{eqnarray}
Note that the operators $ \sigma_{x}^{(j)}, \sigma_{y}^{(j)}$ are not diagonal in the quadrature basis.

{\bf Measurement of the pseudo-spin operators}.
Since the correspondence $ m \>\leftrightarrow \>  (m_1,m_2,..,m_N) $ is unique, an experimental coarse-grained $x-$quadrature measurement 
finding $x\in I_{m,s}$ for some integer $s$ yields the eigen values of  $\sigma_z^{(1)},...\sigma_z^{(N)}$. These are 
homodyne measurements. Measurement of a single operator  $\sigma_z^{(j)}$ is an even more coarse-grained quadrature measurement.
For example, the projection to $\sigma_z^{(j)}=1$ of a state $|\psi> $ just removes those regions of $x=aL(s-1/2)+am+y$ 
which correspond to $m_j=-1$,irrespective of the values of $m_k, k\neq j$,
\begin{equation}
\frac {1+\sigma_{z}^{(j)}}{2} |\psi>=\sum_{s,m,m_j\neq -1} \int_{y=0}^{a} dy \> |m,s,y><m,s,y|\psi>. \nonumber
\end{equation}

Towards the 
measurement of $\sigma_x^{(1)},...\sigma_x^{(N)}$ and $\sigma_y^{(1)},...\sigma_y^{(N)}$, it is sufficient to be able to realize their 
eigen states.The eigen value equations,

\begin{eqnarray}
 && (\sigma_y^{(j)}\mp 1) (|m,s,y> \pm i m_j|m - 2^{(j-1)}m_j,s,y>)=0 ,\nonumber\\
&& (\sigma_x^{(j)}\mp 1)(|m,s,y> \pm |m - 2^{(j-1)}m_j,s,y> )=0.
\end{eqnarray} 
show that the corresponding eigen states are just superpositions of a state with one value of quadrature and another state 
suitably phase-shifted and with the quadrature translated by $ 2^{(j-1)}m_j a $. As stated by Braunstein 
and van Loock \cite{Braunstein-Loock}, quadrature translations are easy and efficient using continuous variable techniques.  
So, a beam-splitter may split the beam into two beams, apply a quadrature translation on one of them and recombine  the two beams 
with their paths being adjusted to obtain the required phase difference. Though less easy than a measurement 
of $\sigma_z^{(j)}$,  measurements of $\sigma_x^{(j)}$, and $\sigma_y^{(j)}$ (and analogously a component of $\vec{\sigma}^{(j)} $ along any direction) 
seem possible, and potentially rewarding .  

{\bf Post-measurement states}. We see from above that a pure state $|\psi>$ corresponds to eigen value of  $\sigma_x^{(1)}=\pm 1 $ if ,
for all $m_2,..,m_N,s,y$,
\begin{eqnarray}
&& <m_1=1,m_2,..,m_N,s,y|\psi>\nonumber\\
&& = \pm <m_1=-1,m_2,..,m_N ,s,y|\psi>\>\>.
\end{eqnarray}
This condition remains unaffected by projection to $\sigma_z^{(j)}=\pm 1 $, if $ j\neq 1$ , because the projection just removes the $m_j=\mp 1$ 
component of the state; i.e. the eigen value of $\sigma_x^{(1)} $ is unaltered by a measurement of $\sigma_z^{(j)}, j\neq 1$. If $N=2,j=2$,
the commutation of $\sigma_z^{(2)}$ and $\sigma_x^{(1)}$ is clearly exhibited by the equation,
\begin{eqnarray}
 \sigma_z^{(2)}\sigma_x^{(1)}=\sigma_x^{(1)}\sigma_z^{(2)}\nonumber \\
  =\sum_{s,m_1,m_2} \int_{y=0}^{a} dy \>m_2 |m_1,m_2,s,y><-m_1,m_2,s,y|.
\end{eqnarray}
This enables the proof that for any arbitrary state with density operator $\rho$ , if $\sigma_z^{(2)}=A$ is measured first to 
obtain the state $\rho'$,a subsequent measurement of $\sigma_x^{(1)}=B$ on the post-measurement state must yield the same expectation value 
as in the initial state. If two self-adjoint operators $A$ and $B$ commute, so do their projectors $P_A^l$ and $P_B^k$.
On measuring $A$, 
\begin{equation}
 \rho \rightarrow \rho '=\sum_l P_A^l \rho P_A^l.
\end{equation}
A measurement of $B$ on the post-measurement state yields the  expectation value ,
\begin{eqnarray}
 Tr \rho '\>B= Tr \sum_l P_A^l \>\rho \> P_A^l B= Tr \sum_l P_A^l \>\rho\> B P_A^l \nonumber\\
 =Tr \rho \>B \sum_l P_A^l P_A^l=Tr \rho \>B,
\end{eqnarray}
which is the same as the expectation value in the initial state. Here we used the commutation of $ P_A^l $ with $B$, 
the cyclicity of the Trace, the idempotence and the completeness of the projectors $P_A^l $. Though the pseudo-spin 
operators $\sigma_x^{(1)}$, and $\sigma_z^{(2)}$ refer to the same continuous variable system, their joint measurement 
characteristics are parallel to those of two independent spin-half systems.

 {\bf Phase space displaced pseudo-spin operators}. The unitary operator for phase space displacement is,
 \begin{equation}
  D(\alpha)=\exp{(i\hat{x}\bar {p}-i\hat{p}\bar{q})},\>\alpha \equiv ( \bar{q}+i\bar{p})/\sqrt{2},
 \end{equation}
 in units  $\hbar=1 $, where the quadrature operators  $\hat{x}$, $\hat{p}$  obey $[\hat{x},\hat{p}]=i $. Then,
 \begin{eqnarray}
   D(\alpha)|x>&=& \exp{(ix\bar {p}+i \frac{\bar{p} \bar{q}}{2})} |x+\bar{q}> \\
   <x|D(\alpha)|\psi>&=&\exp{(ix\bar {p}-i \frac{\bar{p} \bar{q}}{2})} <x-\bar{q}|\psi> .
   \end{eqnarray}
For an arbitrary operator $A$, and state $|\psi>$, I define the displaced operator $A_\alpha $ and displaced state $|\psi_\alpha >$,
\begin{equation} 
 A_\alpha =D(\alpha) \> A \> D(\alpha)^\dagger, \>\> |\psi_\alpha >= D(\alpha)|\psi>\>.
\end{equation}

 Hence, the displaced pseudo-spin operators are given by,

\begin{eqnarray}
&& ( \sigma_{x,\alpha}^{(j)}\pm i \sigma_{y,\alpha}^{(j)})= \exp{(\pm ia\>2^{j-1 }\bar{p} ) }\sum_{s,m} (1 \mp m_j)\nonumber\\
&& \times \int_{y=0}^{a} dy|m \pm 2^{j-1},s,y+\bar{q}><m,s,y+\bar{q}|,
\end{eqnarray}
and 
\begin{equation}
 \sigma_{z,\alpha}^{(j)} =\sum_{s,m} \int_{y=0}^{a} dy \>m_j |m,s,y+\bar{q}><m,s,y+\bar{q}|,
\end{equation}
where,  $s$ is summed over all integers and  
$m$ over all integers $\in [0,2^N -1] $. 

{\bf Bell correlations}. I now define Bell correlations of the pseudo-spin operators and phase space displaced pseudo-spin operators . 
If $\vec{a}^{(j)}$ are unit vectors,
\begin{eqnarray}
& & (\vec{\sigma}^{(j)}.\vec{a}^{(j)})^2=1,\nonumber\\
& & [\vec{\sigma}^{(j)}.\vec{a}^{(j)},\vec{\sigma}^{(j')}.\vec{a}^{(j')}]=0,\>j'\neq j.
\end{eqnarray}
The observables
\begin{equation}\label{A^i}
 A^{(j)}(a^{(j)})\equiv(\vec{\sigma}^{(j)}.\vec{a}^{(j)})\equiv A^{(j)};
\end{equation}
and 
\begin{equation}\label{A^i'}
 A^{(j)}(a^{(j)})'\equiv(\vec{\sigma}^{(j)}.(\vec{a}^{(j)})')\equiv ( A^{(j)})';
\end{equation}
have eigenvalues $\pm 1$. 
For brevity we may sometimes  write $A^{(j)}$ and  $(A^{(j)})'$ instead of $A^{(j)}(a^{(j)})$, and $A^{(j)}(a^{(j)})'$, 
but it will be understood that $A^{(j)}$ depends only on  $a^{(j)} $ and $(A^{(j)})'$ depends only on $(a^{(j)})' $ .
 $A^{(j)}$ commutes with $(A^{(k)})$ and  $(A^{(k)})'$  for  $j\neq k$. Consider the quantum operators, 
 
\begin{eqnarray}
 &&  E^{(N)}(a^{(1)},a^{(2)},..,a^{(N)},(a^{(1)})',(a^{(2)})',..,(a^{(N)})')\nonumber\\
 && = \prod_{j=1}^N \big(A^{(j)}(a^{(j)})+iA^{(j)}(a^{(j)})'\big) ,
  \end{eqnarray}
or, suppressing dependences on the orientations $ a^{(j)},( a^{(j)})'$ ,
\begin{equation}\label{E^{(N)}}
  E^{(N)}\equiv \prod_{j=1}^N \big(A^{(j)}+i(A^{(j)})'\big) \equiv E_1^{(N)} +iE_2^{(N)};
\end{equation}
where the Hermitian operators $  E_1^{(N)} ,E_2^{(N)}$ 
 \begin{eqnarray}\label{E1,E2}
   E_1^{(N)}\equiv (E^{(N)} +(E^{(N)})^\dagger)/2;\nonumber\\
   E_2^{(N)}\equiv (E^{(N)} -(E^{(N)})^\dagger)/(2i),
 \end{eqnarray}
are linear combinations of $2^{(N-1)}$ terms ,each term being a product of $N$ commuting observables and hence 
experimentally measurable . Since ,
\begin{equation}
   E^{(N+1)} =   ( E_1^{(N)} +iE_2^{(N)}) (A^{(N+1)}+i(A^{(N+1)})'),\label{RecursionE}
\end{equation}
we can express higher order Bell operators in terms of lower order ones, as in (\cite{roy-singh1991},\cite{Werner},\cite{Chen-Zhang} ).
I define conveniently normalized even and odd order Hermitian Bell operators,

\begin{eqnarray}\label{Bell-operators}
&& B_1^{(2r)}=\frac{E_1^{(2r)} +E_2^{(2r)} }{2^r };\>B_2^{(2r)}=\frac{E_1^{(2r)} -E_2^{(2r)} }{2^r }\>,r=1,2,..\nonumber\\
&& B_1^{(2r+1)}=\frac{E_1^{(2r+1)} }{ 2^r}; \> B_2^{(2r+1)}=\frac{E_2^{(2r+1)} }{2^r}\>,r=0,1,2,..
\end{eqnarray}
 I then have the Bell operator recursion relations,
\begin{eqnarray}\label{For Cirel'son1}
  B_1^{(2r)}=B_1^{(2r-1)}\frac{A^{(2r)}+(A^{(2r)})'}{2}&&\nonumber\\
  +B_2^{(2r-1)}\frac{A^{(2r)}-(A^{(2r)})'}{2} \>,&&\\
  B_2^{(2r)}= -B_2^{(2r-1)}\frac{A^{(2r)}+(A^{(2r)})'}{2} && \nonumber\\\label{For Cirel'son2}
  +B_1^{(2r-1)}\frac{A^{(2r)}-(A^{(2r)})'}{2}\>,&&
\end{eqnarray}
or equivalently,
\begin{eqnarray}\label{For Cirel'son'}
  B_1^{(2r+1)}=B_2^{(2r)}\frac{A^{(2r+1)}+(A^{(2r+1)})'}{2}&&\nonumber\\
  +B_1^{(2r)}\frac{A^{(2r+1)}-(A^{(2r+1)})'}{2} ,&&
\end{eqnarray}
  \begin{eqnarray}\label{Recursion}
  B_2^{(2r+1)}=B_1^{(2r)}\frac{A^{(2r+1)}+(A^{(2r+1)})'}{2}&&\nonumber\\ 
  -B_2^{(2r)}\frac{A^{(2r+1)}-(A^{(2r+1)})'}{2} .&&
\end{eqnarray}
Similarly,the corresponding displaced observables 
$$ A^{(j)}(a^{(j)})_\alpha = D(\alpha) \> A^{(j)}(a^{(j)})\> D(\alpha)^\dagger,$$
have eigenvalues $\pm 1$ and are mutually commuting for different values of $j$. Defining,
\begin{equation}
  E_\alpha^{(N)}\equiv \prod_{j=1}^N \big(A^{(j)}_\alpha+i(A^{(j)})'_\alpha \big) \equiv E_{1,\alpha}^{(N)} +iE_{2,\alpha}^{(N)},
\end{equation}
displaced analogues of Eqns. ( \ref{E1,E2})-(\ref{Recursion} ) are obtained by the replacements:
\begin{eqnarray}
 A^{(j)}\rightarrow A^{(j)}_\alpha \>, (A^{(j)})' \rightarrow (A^{(j)})'_\alpha \>, E^{(N)}\rightarrow E_\alpha^{(N)}, \nonumber\\
  E_i^{(N)}\rightarrow E_{i,\alpha}^{(N)}\>,B_i^{(N)}\rightarrow B_{i,\alpha}^{(N)}.
\end{eqnarray}

The quantum Bell correlations are given by the  expectation values,
\begin{eqnarray}\label{QM_correlations}
 &&<B_i^{(N)}>_{QM}= Tr \rho \> B_i^{(N)},\nonumber\\
 &&< B_{i,\alpha}^{(N)}>_{QM}= Tr \rho \> B_{i,\alpha}^{(N)}\>,
\end{eqnarray}
where $i=1,2  $,and  $\rho$ is the density operator for the state. 

We want to compare the quantum Bell correlations with the predictions of a non-contextual hidden variable theory.

{\bf Non-contextual hidden variables (NCHV) }. In a non-contextual stochastic hidden variable theory,  the 
state with hidden variables $\lambda$ with probability distribution $\mu (\lambda)$, specifies the values or at 
least expectation values corresponding to the $j$-th observables (\ref{A^i}) as $A^{(j)}(\lambda,a^{(j)}) $ which must lie in the interval
$[-1,+1] $ and  be independent of the orientations of $\vec{a}^{(j')} $ for $j'\neq j$. I denote the $NCHV$ expectation value of a quantum operator 
$A$ by $ <A>=<A>_{NCHV}$, and the corresponding quantum expectation value by $ <A>_{QM}$. Hence the $NCHV$ expectation value 
corresponding to the operator $E^{(N)}$  is given by
 
\begin{eqnarray}
&& <E^{(N)}>=  <E_1^{(N)}> +i<E_2^{(N)}>\nonumber\\
 && =\int d\lambda  \mu(\lambda)\prod_{j=1}^N \big(A^{(j)}(\lambda)+i(A^{(j)})'(\lambda)\big)\nonumber\\
 && \equiv \int d\lambda  \mu(\lambda)E^{(N)}(\lambda), \\
 &&E^{(N)}(\lambda)=E_1^{(N)}(\lambda)+iE_2^{(N)}(\lambda),
 \end{eqnarray}
 where $E_1^{(N)}(\lambda),E_2^{(N)}(\lambda) $ are the real and imaginary parts of $E^{(N)}(\lambda) $. It will be understood 
 that $A^{(j)}(\lambda)=A^{(j)}(\lambda,a^{(j)})$ also depends  on  $a^{(j)} $ ,and 
 $(A^{(j)})'(\lambda)=(A^{(j)})'(\lambda,(a^{(j)})')$ depends also on $(a^{(j)})' $ .The normalisation conditions are,
 
 \begin{eqnarray} \label{normalisation condition}
 && \int d\lambda \mu(\lambda) =1,\>; \mu(\lambda) \geq 0,\nonumber\\
 && |A^{(j)}(\lambda) | \leq 1\>; |(A^{(j)})'(\lambda)| \leq 1\>.
 \end{eqnarray}
 They imply already that,
 \begin{eqnarray}
 && |E_i^{(1)}(\lambda)| \leq 1,\>; |<E_i^{(1)}> | \leq 1; \>i=1,2\>; \\
 && |E^{(N)}(\lambda)| \leq 2^{N/2}\>; |<E^{(N)}> | \leq 2^{N/2},
 \end{eqnarray}
but the $N > 1 $ results can be improved.

 The NCHV expectation value $<E_\alpha^{(N)}>$ is
 \begin{eqnarray}
 && <E_\alpha^{(N)}>=  <E_{1,\alpha}^{(N)}> +i<E_{2,\alpha}^{(N)}> \nonumber\\
 && =\int d\lambda  \mu(\lambda)\prod_{j=1}^N \big(A^{(j)}_\alpha(\lambda)+i(A^{(j)})'_\alpha(\lambda)\big)\nonumber\\
 && \equiv \int d\lambda  \mu(\lambda)E_\alpha^{(N)}(\lambda) ,\\
 &&E_{\alpha}^{(N)}(\lambda)= E_{1,\alpha}^{(N)}(\lambda)+i E_{2,\alpha}^{(N)}(\lambda),
  \end{eqnarray}
where,$A^{(j)}_\alpha(\lambda)=A^{(j)}_\alpha(\lambda,a^{(j)})$ depends also on  $a^{(j)} $ ,and 
 $(A^{(j)})'_\alpha(\lambda)=(A^{(j)})'_\alpha(\lambda,(a^{(j)})')$ depends also on $(a^{(j)})' $ . Further,
 
\begin{equation}
 |A^{(j)}_\alpha(\lambda) | \leq 1 \>; |(A^{(j)})'_\alpha(\lambda)|\leq 1,
\end{equation}
and hence,
\begin{eqnarray}
  && |E_{i,\alpha}^{(1)}(\lambda)| \leq 1,\>; |<E_{i,\alpha}^{(1)}> | \leq 1; \>i=1,2\>; \\
 && |E_\alpha^{(N)}(\lambda)| \leq 2^{N/2}\>; |<E_\alpha^{(N)}>| \leq 2^{N/2},
 \end{eqnarray}
which too can be improved for $N>1$.

 Writing the NCHV value of each operator $A$ in the hidden state $\lambda$ as $A(\lambda)$,
 the NCHV correlations corresponding to the Bell operators $B_i^{(j)}$ , $B_{i,\alpha}^{(j)} $ are given by,
 \begin{eqnarray}
 && <B_i^{(j)}>= \int d\lambda  \mu(\lambda) B_i^{(j)}(\lambda)\>; \nonumber\\
 && <B_{i,\alpha}^{(j)}>= \int d\lambda  \mu(\lambda) B_{i,\alpha}^{(j)}(\lambda)\>;i=1,2
 \end{eqnarray}
  Thus, the operator relations 
 (\ref{RecursionE})-(\ref{Recursion}) yield corresponding recursion relations between the hidden variable values. E.g.
 
 \begin{eqnarray}\label{RecursionE'}
  && E^{(N+1)}(\lambda) =   ( E_1^{(N)} +iE_2^{(N)})(\lambda)\nonumber\\
  && \times (A^{(N+1)}+i(A^{(N+1)})')(\lambda),
\end{eqnarray},
\begin{eqnarray}\label{Bell-operators'}
&& B_1^{(2r)}(\lambda)=\frac{E_1^{(2r)}(\lambda) +E_2^{(2r)}(\lambda) }{2^r };\>r=1,2,..\nonumber\\
&&B_2^{(2r)}(\lambda)=\frac{E_1^{(2r)}(\lambda) -E_2^{(2r)}(\lambda) }{2^r }\>,r=1,2,..\nonumber\\
&& B_1^{(2r+1)}(\lambda)=\frac{E_1^{(2r+1)}(\lambda) }{ 2^r}; \>r=0,1,2,..\nonumber\\
&& B_2^{(2r+1)}(\lambda)=\frac{E_2^{(2r+1)}(\lambda) }{2^r}\>,r=0,1,2,..
\end{eqnarray}
 These lead to  recursion relations for hidden variable values of the Bell operators,
 \begin{eqnarray}\label{Recursion-even}
  B_1^{(2r)}(\lambda)=B_1^{(2r-1)}(\lambda)\frac{(A^{(2r)}+(A^{(2r)})')(\lambda)}{2}&&\nonumber\\
  +B_2^{(2r-1)}(\lambda)\frac{(A^{(2r)}-(A^{(2r)})'(\lambda)}{2} \>,&&\\
  B_2^{(2r)}(\lambda)= -B_2^{(2r-1)}(\lambda)\frac{(A^{(2r)}+(A^{(2r)})')(\lambda)}{2} && \label{Recursion-even'}\nonumber\\
  +B_1^{(2r-1)}(\lambda)\frac{(A^{(2r)}-(A^{(2r)})')(\lambda)}{2}\>,&&
\end{eqnarray}
or equivalently,
\begin{eqnarray}\label{Recursion-odd}
  B_1^{(2r+1)}(\lambda)=B_2^{(2r)}(\lambda)\frac{(A^{(2r+1)}+(A^{(2r+1)})')(\lambda)}{2}&&\nonumber\\
  +B_1^{(2r)}(\lambda)\frac{(A^{(2r+1)}-(A^{(2r+1)})')(\lambda)}{2} ,&&
 \end{eqnarray}
 \begin{eqnarray}\label{Recursion-odd'}
 B_2^{(2r+1)}(\lambda)=B_1^{(2r)}(\lambda)\frac{(A^{(2r+1)}+(A^{(2r+1)})')(\lambda)}{2}&&\nonumber\\
  -B_2^{(2r)}(\lambda)\frac{(A^{(2r+1)}-(A^{(2r+1)})')(\lambda)}{2} .&&
\end{eqnarray}

These relations enable a recursive proof of $N$-qubit $NCHV$ inequalities similar to the original Bell-CHSH locality 
inequalities \cite{Bell1964}, and their $N$-party generalisations  (\cite{mermin1990},\cite{roy-singh1991},\cite{ardehalli1992} ).
Thus, from  Eqns. (\ref{Recursion-even} ),(\ref{Recursion-even'} ),
\begin{eqnarray}\label{recursionproof}
 && |B_i^{(2r)}(\lambda)| \leq   \big (|(A^{(2r)}(\lambda)+(A^{(2r)})'(\lambda))|\nonumber \\
 &&+|(A^{(2r)}-(A^{(2r)})'(\lambda)\big )|/2 \nonumber \\
 && \times max \big ( |B_1^{(2r-1)}(\lambda)|,|B_2^{(2r-1)}(\lambda)|\big )\>,i=1,2.
 \end{eqnarray}
 Using the normalisation conditions (\ref{normalisation condition} ) we have, 
\begin{equation}
 |(A^{(2r)}+(A^{(2r)})')(\lambda)|  
 +|(A^{(2r)}-(A^{(2r)})'(\lambda)| \leq 2\>.
\end{equation}
On multiplying Eqn. (\ref{recursionproof} )by $\mu(\lambda ) $ and  integrating over $\lambda$
we obtain,
\begin{eqnarray}
 && max (|< B_1^{(2r)} > | ,|<  B_2^{(2r)}> | ) \nonumber\\
 && \leq max (|< B_1^{(2r-1)} > | ,|<  B_2^{(2r-1)}> | ) ,
 \end{eqnarray}
 for any positive integer $r$. This proof is  analogous to the original two party Bell-CHSH proof \cite{Bell1964} and 
 an alternative to the variational $N$ party proofs (\cite{mermin1990},\cite{roy-singh1991},\cite{ardehalli1992} ).
 Similarly, Eqns. (\ref{Recursion-odd}),(\ref{Recursion-odd'}) yield,for any positive integer $r$,
\begin{eqnarray}
 && max (|< B_1^{(2r+1)} > | ,|<  B_2^{(2r+1)}> | ) \nonumber\\
 && \leq max (|< B_1^{(2r)} > | ,|<  B_2^{(2r)}> | ).
\end{eqnarray}

From $B_i^{(1)} =E_i^{(1)}$ and $|<E_i^{(1)}> | \leq 1 $ for $i=1,2$, we obtain recursively the $NCHV$ Bell inequalities 
 in a single continuous variable system,

\begin{equation}\label{NCHV1}
 max (|< B_1^{(j)} > | ,|<  B_2^{(j)}> | ) \leq 1 \>; j=1,2,..N\>.
\end{equation}
Exactly the same procedure yields the $NCHV$ Bell inequalities for the displaced operators in  a single continuous variable system,

\begin{equation}\label{NCHV2}
 max (|< B_{1,\alpha}^{(j)} > | ,|<  B_{2,\alpha}^{(j)}> | ) \leq 1 \>; j=1,2,..N\>.
\end{equation}
 
 Our normalisations (\ref{Bell-operators}) of the Bell operators have ensured that their $NCHV$ expectation values are bounded by unity.
 What are the upper limits on the quantum expectation values ?
 
 {\bf Maximum possible quantum violations of NCHV inequalities}.
 The Cirel'son theorem in the two qubit case \cite{Cirel'son} and generalised Cirel'son theorems in $N$-qubit case \cite{Werner} and 
 $N$ continuous variable systems (\cite{Chen}, \cite{Chen-Zhang}) set limits on maximum possible quantum violations of local hidden variable 
 inequalities . I derive analogous limits on quantum violations of $NCHV$ inequalities using pseudo-spin observables for a single 
 continuous variable system.
 
 From Eqn. (\ref{For Cirel'son1}), setting $M=2r$ and writing the pseudo-spin operators explicitly, I obtain,
 \begin{eqnarray}
   B_1^{(M)}=B_1^{(M-1)}\frac{\vec{\sigma}^{(M)}.\big(\vec{a}^{(M)}+(\vec{a}^{(M)})'\big)}{2}&&\nonumber\\
  +B_2^{(M-1)}\frac{\vec{\sigma}^{(M)}.\big(\vec{a}^{(M)}-(\vec{a}^{(M)})'\big)}{2},&&
   \end{eqnarray}
we get,
\begin{eqnarray}
   (B_1^{(M)})^2=(B_1^{(M-1)})^2\frac{\big(1+ \vec{a}^{(M)}.(\vec{a}^{(M)})'\big)}{2}&&\nonumber\\
  +(B_2^{(M-1)})^2\frac{\big(1- \vec{a}^{(M)}.(\vec{a}^{(M)})'\big)}{2}&&\nonumber\\
  +\frac{i}{2}[ B_2^{(M-1)},B_1^{(M-1)} ]\>\vec{\sigma}^{(M)}.\big(\vec{a}^{(M)}\times(\vec{a}^{(M)})'\big) &&
   \end{eqnarray}
 Using $ ||A^2||=||A||^2$ for a Hermitian $A$, I have, for $M$ even,
 \begin{equation}\label{bound1}
  ||(B_1^{(M)})^2 || \leq 2 Max \big (||B_1^{(M-1)}||^2 ,||B_2^{(M-1)}||^2 \big).
 \end{equation}
 Similarly, Eqn. (\ref{For Cirel'son2} ) yields, for even $M$,
 \begin{equation}\label{bound2}
  ||(B_2^{(M)})^2 || \leq 2 Max \big (||B_1^{(M-1)}||^2 ,||B_2^{(M-1)}||^2 \big).
 \end{equation}
 Similarly , Eqns. ( \ref{For Cirel'son'}),(\ref{Recursion} ) imply that  the relations  (\ref{bound1}) ,(\ref{bound2} ) also 
 hold for odd $M$. Using $(B_1^{(1)})^2=1,(B_2^{(1)})^2=1 $, I now obtain recursively, both for $N$ even, and for $N$ odd,
 \begin{equation}
  ||(B_1^{(N)})^2 || \leq 2^{N-1}\>; ||(B_2^{(N)})^2 || \leq 2^{N-1}\>.
 \end{equation}
The same bounds follow for the displaced operators,
\begin{equation}
  ||(B_{1,\alpha}^{(N)})^2 || \leq 2^{N-1}\>; ||(B_{2,\alpha}^{(N)})^2 || \leq 2^{N-1}\>.
 \end{equation}
 Hence, for an arbitrary normalzed quantum state , 
\begin{equation}
 |<B_i^{(N)}>_{QM}|  \leq ||B_i^{(N)}||\leq 2^{(N-1)/2 }\>,i=1,2 \>
\end{equation}
and,
\begin{equation}
 |<B_{i,\alpha}^{(N)}>_{QM}|  \leq ||B_{i,\alpha}^{(N)}||\leq 2^{(N-1)/2 }\>,i=1,2 \>.
\end{equation}

We see that just as for the Mermin-Roy-Singh multiparty locality inequalities (\cite{mermin1990},\cite{roy-singh1991},\cite{ardehalli1992} ), 
quantum correlations can violate the $NCHV$ limits for $N$ pseudo spins in a continuous variable system    
by at most a factor $ 2^{(N-1)/2}$. Thus the generalised Cirel'son theorem  for the $NCHV$ single system case is similar to that for  
locality inequalities for the $N$ system case (\cite{Cirel'son},  \cite{Werner} ,\cite{Chen}, \cite{Chen-Zhang}). Can this maximal violation 
be reached?

{\bf Quantum state showing maximal violation of NCHV inequality for a given N}. Consider the special choice,
\begin{equation}\label{E_0}
 E^{ (N)} \equiv  \prod_{j=1 }^N (\sigma_x^{(j)}- i \sigma_y^{(j)}) ,
\end{equation}
where, $\sigma_x^{(j)} ,\sigma_y^{(j)} $ are the pseudo-spin operators ;  the 
Hermitian operators $E_i^{(N)},B_i^{(N)}$ for $i=1,2$ are derived from it using Eqns. 
(\ref{E1,E2}) and (\ref{Bell-operators}). I evaluate the quantum Bell correlations 
for the particular state,
\begin{eqnarray}
&& |\psi_0> =\sum_{s=-\infty}^\infty \sum_{m=0,1,..}^{(2^N-1 ) }\int_{y=0}^{a} dy\nonumber\\
&&|m,s,y> \psi_0 (aL(s-\frac{1}{2})+am+y ).
\end{eqnarray}
where $ \psi_0 (aL(s-\frac{1}{2})+am+y )= 0$, unless $m = 0 \> or \> L-1,$ ,i.e.

\begin{eqnarray}\label{m=0,L-1}
&& |\psi_0>=\sum_s\int_0^a dy \big (|0,s,y> \psi_0 (aL(s-\frac{1}{2})+y ) \nonumber\\
&& +|L-1,s,y> \psi_0 (aL(s-\frac{1}{2})+a(L-1)+y) \big ),
 \end{eqnarray}
 with the normalisation condition, $<\psi_0|\psi_0>=1$. E.g., the  particular choice  
$$\psi_0 (aL(s-1/2) +y)=(2/(\pi (2s-1) ) \chi (y),$$
$$\>\int_0^a dy |\chi (y)|^2=1/2 \>,$$ 
 obeys the normalization condition  because (see\cite{Gradshteyn}) $$ \sum_{k=1}^{\infty}(2k-1)^{-2}= \pi^2/8 $$ .

 The definition (\ref{E_0}) of $E^{(N)}$  and of the $ \sigma$ matrices  yield,
   \begin{eqnarray}
&& <\psi_0|E^{(N)}|\psi_0>= 2^N \sum_s\int_0^a dy \>\psi_0 ^* (aL(s-1/2)+y)\nonumber\\
&& \times \psi_0 (aL(s-1/2)+a(L-1)+y). \nonumber
 \end{eqnarray}
I now choose,
\begin{eqnarray}
 &&\psi_0 (aL(s-1/2)+a(L-1)+y) \nonumber\\
 && = \exp {(i\theta) }\psi_0 (aL(s-1/2)+y),
\end{eqnarray}
and use the normalization condition to obtain,
\begin{equation}
 <\psi_0|E^{(N)} |\psi_0>= 2^{N-1} \exp {(i\theta) }.
\end{equation}
This gives the quantum correlations,
\begin{eqnarray}
N\>odd: |<B_1^{(N)}>_{QM}|&=&2^{(N-1)/2},\>\theta=0 ; \nonumber\\
|<B_2^{(N)}>_{QM}|&=&2^{(N-1)/2},\>\theta=\pi/2 ;
\end{eqnarray}
\begin{eqnarray}
N \>even: |<B_1^{(N)}>_{QM}|=2^{(N-1)/2},\>\theta= \pi/4\>;&&\nonumber\\
|< B_2^{(N)}>_{QM}|=2^{(N-1)/2},\>\theta=-\pi/4 &&
\end{eqnarray}
Choosing $\theta=0$,or $\pi/2$ for $N$ odd, and $\theta=\pm \pi/4$ for $N$ even, we see that the NCHV inequalities (\ref{NCHV1}) 
are violated by quantum mechanics by the maximal factor $2^{(N-1)/2 }$ which grows exponentially with the chosen $N$. 
This concludes the formal proof, but the state may not be easily realizable. I now show that nearly maximal violation is possible 
using  very familiar states.

{\bf Non-contextuality violation by the single mode squeezed vacuum (SMSV) state }. I  now consider the SMSV wave function ,  
which is easy to realise quantum optically, and has a positive definite Wigner function,
\begin{equation}
\psi (x)= C \exp {(-x^2 /(4 \sigma ^2) ) }  ;\>C= ( 2\pi \sigma ^2)^{-1/4 }\>. 
\end{equation}
The relation with the squeezing parameter $r'$ is given by $\sigma ^2 = \frac{1}{2} \exp{(-2r' ) }$,
where $r' > 0$ corresponds to position squeezing, and $r'<0$  to momentum squeezing.
Rewriting the states $|x>$ in terms of the states $|m,s,y>$ with definite eigenvalues of $\sigma_z^{(j)} $, we have,
 \begin{eqnarray}\label{SMSV}
&& |\psi> =\sum_{s=-\infty}^\infty \sum_{m=0,1,..}^{(2^N-1 ) }\int_{y=0}^{a} dy |m,s,y>\nonumber\\
&& \times C \exp {\big(-(aL(s-1/2)+am+y )^2  /(4 \sigma ^2) \big ) } .
\end{eqnarray}
Note that, $ m= \frac{L}{2}-1 \leftrightarrow x-Las\in (-a,0) $ and $ m= \frac{L}{2}\leftrightarrow x-Las\in (0,a)$.
Further, $$m=\frac{L}{2}-1 \leftrightarrow  (m_1,..,m_N)=(1,..,1,-1) $$ , and 
 $$m=\frac{L}{2} \leftrightarrow (m_1,..,m_N)=(-1,..,-1,1) $$. This time we choose 
   the operator $E^{(N)}$ defined by
   \begin{equation}\label{E'}
  E^{(N)} \equiv  (\sigma_x^{(N)}- i \sigma_y^{(N)})\prod_{j=1 }^{(N-1)} (\sigma_x^{(j)}+ i \sigma_y^{(j)}) 
 \end{equation}
 which takes states with  $m=L/2$, to states with $m=L/2-1$, 
\begin{equation}
 E^{(N)}|m=L/2, s,y>= 2^N |m=L/2-1,s,y>;
\end{equation}
 and annihilates all states with $m\neq \frac{L}{2} $.
Using this ,and the orthonormality of the states $|m,s,y> $ I obtain, for odd $N$,
\begin{eqnarray}
&& <\psi| E^{(N)} |\psi> =\sum_{s=-\infty}^\infty \int_{y=0}^{a} dy\> \nonumber\\
&& \times 2^N C^2 \exp {\big(-\frac{(aLs-a +y )^2 +(aLs+y )^2 } {4 \sigma ^2 } \big ) } ,
\end{eqnarray}
and
\begin{equation}
<B_1^{(N)}>_{QM}= Re  \frac{ <\psi| (E^{(N)}) |\psi> } { 2^{(N-1)/2} }.
\end{equation}
 Since each $s$ gives a positive contribution to $ <B_1^{(N)}>_{QM}$, I obtain for odd $N$, a lower bound on this Bell 
 correlation by keeping only $s=0$,
 \begin{eqnarray}
  && <B_1^{(N)}>_{QM}\>\ge 2^{(N+1)/2} \exp{(-\frac{\mu^2}{2}) }A(\mu), \nonumber\\
  &&\mu \equiv \frac {a}{2\sigma},\>A(\mu)\equiv \frac{1}{\sqrt{2 \pi}}\int_{-\mu}^{\mu} dt\> exp {(-\frac{t^2}{2} ) }.
 \end{eqnarray}
The choice $ a=1.8 \sigma $ yields,
\begin{equation}\label{B_1QM}
 <B_1^{(N)}>_{QM} \ge 0.842 \times 2^{(N-1)/2 }\>; N\> odd \>.
\end{equation}
This contradicts the $NCHV$ bound $|<B_1^{(N)}>|\leq 1 $ 
by a factor $0.842 (\sqrt{2} )^{N-1} $ for odd $N \geq 3$ .This violation holds for 
the vacuum state ($\sigma^2 =1/2$) too, with the choice $a=1.8/\sqrt{2}$. 
  
  {\bf Non-contextuality violation by a single mode squeezed coherent state }. Consider a squeezed coherent state $|\psi_\alpha>$ that is 
  obtained by displacing the squeezed vacuum state $|\psi>$ of the last section,
  $$|\psi_\alpha>= D(\alpha) |\psi > $$. It has a positive Wigner function,
  \begin{equation}
   W_\alpha (x,p)= \frac{1}{\pi} \exp{\big( -\frac{(x-\bar{x})^2 } {2 \sigma ^2 }-2\sigma^2 (p-\bar{p})^2 \big) }.
  \end{equation}

  Without any extra work, the quantum prediction for expectation value of $  E_\alpha^{(N)} $ obtained by displacement of 
  the operator $E^{(N)}$  ( Eqn. (\ref{E'})) in the  displaced state $ |\psi_\alpha>$ is obtained simply by using,
  \begin{equation}
    <\psi_\alpha |E_\alpha ^{(N)} |\psi_\alpha>= \>  <\psi| E^{(N)} |\psi >\>.
   \end{equation}
   Hence using the result of the last section, for $N$ odd,
   \begin{equation}
  <\psi_\alpha |B_{1,\alpha}^{(N)}|\psi_\alpha>=<\psi|B_1^{(N)}|\psi>\>\ge 0.842 \times 2^{(N-1)/2 }\>, 
  \end{equation}
which again contradicts the $NCHV$ bound on $|<B_{1,\alpha}^{(N)}>| $ 
by a factor $0.842 (\sqrt{2} )^{N-1} $ for $ a=1.8 \sigma $ and odd $N \geq 3$ . This 
violation is obtained for the usual coherent state ( no squeezing ) too, if we choose $a=1.8/\sqrt{2}$, 
showing the inequivalence of non-contextuality with the alternate classicality criterion of `coherence'.
The non-contextuality violations can be tested experimentally.

{\bf Conclusions and comparisons with earlier work}. Non-contexuality is a hallmark of classical behaviour, and 
contextuality a valuable quantum resource \cite{resource}. 
It is shown that arbitrarily large violation of non-contextuality  is  possible even in states of a single continuous variable system with 
positive Wigner function. Although the pseudo-spin observables  are defined here in $x$-space , they are non- diagonal .
It is known that a classical probability description for any diagonal operator such as $|x><x|$ or  $|p><p|$ (or $|c><c|$ where $c$ is an 
eigenvalue of any linear combination of the two quadrature operators), is possible for 
a quantum state with a positive Wigner function. The present work which extends to `contextuality' the earlier work  
(\cite{Banaszek},\cite{Chen},\cite{Chen-Zhang}) on `non-locality' shows very clearly that the 
`classicality' of states with positive  Wigner function does not hold for pseudo-spin operators which are non-diagonal in $x$- space; 
the non-classicality can be arbitrarily large.

There is at least another class of non-contextuality inequalities violated by quantum states irrespective of the 
positivity of the Wigner function. Plastino and Cabello , and others \cite {Cabello} obtained state-independent 
non-contextuality inequalities for continuous (and discrete) variables. In particular state independent inequalities 
involving 18 modular observables for a two dimensional configuration space were shown to be violated by any quantum state 
by a factor $2/\sqrt{3}$.

In contrast with all earlier work, the present non-contextuality inequalities for a one-dimensional configuration space 
can be violated by some quantum states by arbitrarily large factors (when $N$ is chosen large enough).
 Practical  demonstrations of  `contextuality' using the inequalities presented here will involve measurement of 
 Bell correlations of displaced pseudo-spin observables in quantum optical states. The squeezed coherent state seems particularly promising.
 Efficient techniques have been developed for manipulation and measurement of continuous quadratures and their phase space 
 displacements \cite{Braunstein-Loock}. If practical techniques to measure the pseudo-spin observables 
 non-diagonal in quadrature space are developed, applications of large classicality violation to quantum information tasks 
 would be possible. It will be interesting to know if decoherence effects may be less severe in  a single system used 
 here than in  multiparty systems used before.
 
 {\bf Acknowledgements}.  I thank Sibasish Ghosh, Aditi Sen De, Saptarshi Roy, N. Mukunda and Anupam Garg 
for intense discussions during the QFF2018 conference at the Raman Research Institute, Bangalore, April 30 to May 4, 2018,  
Guy Auberson for the reference to phase space Bell inequalities optimal for experimental tests \cite{Grangier}, and A. Asadian 
for references  to their work \cite{Asadian} on applications of modular observables. I thank 
the organisers Urbasi Sinha, Dipankar Home and  Alexandre Matzkin for invitation to this conference, and the Indian National Science 
Academy for the INSA honorary scientist position.


\begin{thebibliography}{99}

\bibitem{Banaszek}
K. Banaszek and K. Wódkiewicz, Phys. Rev. A 58, 4345
(1998); Phys. Rev. Lett. 82, 2009 (1999); Acta Phys. Slo-
vaca 49, 491 (1999); P. van Loock and S. L. Braunstein, Phys. Rev. A 63,
022106 (2001) and quant-ph/0006029;
J. Martin and V. Vennin, arxiv.org/pdf/1605.02944.pdf, and Phys.Rev. A93 (2016) no.6, 062117; 
J.-A. Larsson, Phys. Rev. A70, 022102 (2004), quant-ph/0310140, and  Phys. Rev. A 67, 022108 (2003), quant-ph/0208125.
 C. Vitelli, M. T. Cunha, N. Spagnolo, F. De Martini,and F. Sciarrino, Phys. Rev. A 85, 012104 (2012).

\bibitem{Chen}
Zeng-Bing Chen,Jian-Wei Pan,Guang Hou, and Yong-De Zhang,
Phys. Rev. Lett. {\bf 88},040406-1 (2002);Saptarshi Roy,
Titas Chanda, Tamoghna Das, Aditi Sen De, Ujjwal Sen, arxiv.org/abs/1807.03158.


\bibitem{Chen-Zhang}
Zeng-Bing Chen and Yong-De Zhang,
Phys. Rev. A 65, 044102 (2002) and arXiv:quant-ph/0103082v2 (2002) ;
C. Brukner, M. S. Kim, J.-W. Pan, and A. Zeilinger,
Phys. Rev. A 68, 062105 (2003), quant-ph/0208116.

\bibitem{Gleason} A.M. Gleason,  {\it J. Math. \& Mech.} \underbar{6}, 885
(1957); S. Kochen and E.P. Specker, {\it J. Math. \& Mech.}
\underbar{17}, 59 (1967) ; A. Peres, Phys. Lett. A {\bf 151},107 (1990) and J. Phys. A
{\bf 24}, L175 (1991) ; N. D. Merrnin, Phys. Rev. Lett. {\bf 65}, 3373 (1990) and Rev. Mod. Phys.
{\bf 65}, 803 (1993);A. Martin and S.M. Roy, {\it Phys. Lett.} \underbar{B350},66 (1995); 

\bibitem{Bell1966} 
J. S. Bell, Rev. Mod. Phys. {\bf 38},447(1966).

\bibitem{Einstein1933} A. Einstein, {\it Phiolosopher Scientist}, P.A. Schilp Ed.. Library 
of Living Philosophers, Evanston, Ill. (1949). See esp. Einstein's `autobiographical notes' 
and replies by N. Bohr here and in N. Bohr, {\it Phys. Rev.} {\bf 48} (1935) 696. See also, 
reprints of related articles in ,{\it Quantum Theory and Measurement}, J. A. Wheeler and 
W. H. Zurek Eds., Princeton (1983). 

\bibitem{Bell1964} J. S. Bell, {\it Physics} {\bf 1} (1964) 195; A. Einstein, B. Podolsky and N. Rosen, {\it Phys. Rev.} {\bf 47} (1935) 777;
J. F. Clauser, M. A. Horne, A. Shimony, and R. A. Holt, {\it Phys. Rev. Lett.} {\bf 26} (1969) 880;
S. M. Roy and V. Singh, J. Phys. {\bf A11}, L167 (1978).

\bibitem{resource}
See e.g.   
 R. W. Spekkens, Phys. Rev. {\bf A 71}, 052108 (2005); E. Knill, Nature {\bf 434}, 39 (2005); 
 R. W. Spekkens, D. H. Buzacott, A. J. Keehn, B. Toner,
and G. J. Pryde,Phys. Rev. Lett.{\bf 102},010401 (2009); C. Liang, R. W. Spekkens, and H. M. Wiseman,
 Phys. Rep. {\bf 506}, 1 (2011);
M. Waegell and P. K. Aravind, Phys. Rev. {\bf A88}, 012102 (2013) and 
Phys. {\bf A45}, 405301 (2012); M. Howard, J. Wallman, V. Veitch, and J. Emerson,
“Contextuality supplies the ’magic’ for quantum computation,”, Nature {\bf 510}, 351 (2014); 
Xiao-Dong Yu, Yan-Qing Guo and D. M. Tong, arXiv:1505.026032v2 [quant-ph];J. Bermejo-Vega, N. Delfosse, D. E. Browne, C. Okay,
and R. Raussendorf, “Contextuality as a resource for qubit quantum computation,” arXiv:1610.08529 (2016);
 M. D. Mazurek, M. F. Pusey, R. Kunjwal, K. J. Resch,and R. W. Spekkens,``An experimental test of noncon-
textuality without unphysical idealizations,” Nat. Comm.7 (2016), 10.1038/ncomms11780; 
R. Kunjwal and R. W. Spekkens, Phys. Rev. Lett.{\bf115},110403 (2015); Jaskaran Singh, Kishor Bharti, and Arvind,
Phys. Rev. {\bf A 95}, 062333 (2017) and  arXiv:1612.02616v3 [quant-ph]; Anirudh Krishna,Robert W. Spekkens,and Elie Wolfe,
arXiv:1704.01153v2[quant-ph] ,and New J. Phys 19, 123031 (2017). 

\bibitem{Gisin}
 N. Gisin and A. Peres, Phys. Lett. A 162, 15 (1992)

 \bibitem{mermin1990} N. D. Mermin, {\it Phys. Rev. Lett.} {\bf 65} (1990) 1838.

\bibitem{roy-singh1991} S. M. Roy and V. Singh, {\it Phys. Rev. Lett.} {\bf 67} (1991) 2761.

\bibitem{ardehalli1992} M. Ardehalli, {\it Phys. Rev. A} {\bf 46} (1992) 5375; A. V. Belinskii and D. N. Klyshko,
 {\it Phys. Usp.} {\bf 36} (1993) 653; N. Gisin and H. Bechmann-Pasquinucci,{\it Phys. Lett. A} {\bf 246 }(1998) 1.

 \bibitem{Roy-Singh1993}
S. M. Roy and V. Singh, Phys. Rev. {\bf A48},3379(1993) and Vistas in Astronomy {\bf 37},317 (1993).

\bibitem{braunstein}
 S. Lloyd and S. L. Braunstein, Phys. Rev. Lett.82, 1784 (1999);
S. D. Bartlett, B. C. Sanders, S. L. Braunstein and K. Nemoto, Phys. Rev. Lett. 88, 097904 (2002); 
Stephen D. Bartlett and Barry C. Sanders, arXiv:quant-ph/0110039v2 19 Mar 2002;

\bibitem{Braunstein-Loock}
S. L. Braunstein and P. van Loock, Rev. Mod. Phys. {\bf 77}, 513 (2005) and quant-ph/0410100.

\bibitem{Ou}
Ou, Z. Y Pereira, S. F.; Kimble, H. J.; Peng, K. C. .  Phys. Rev. Lett. 68: 3663 (1992) ;A. Kuzmich, I. A. Walmsley, and L. Mandel, Phys. Rev.
Lett. 85, 1349 (2000); L.-M. Duan et al., Phys. Rev. Lett. 84, 2722 (2000) ; P. van Loock and S. L. Braunstein, Phys. Rev. Lett. 84,
3482 (2000); Villar, A. S.; Cruz, L. S.; Cassemiro, K. N.; Martinelli, M.;
Nussenzveig, P. . 
Phys. Rev. Lett. 95: 243603 (2005), arXiv:quant-ph/0506139;
 T. Yarnall, A. F. Abouraddy, B. E. A. Saleh, and M. C.Teich, Phys. Rev. Lett. 99, 170408 (2007);
M. M. Dorantes and J. L. Lucio M, Journal of Physics A, Mathematical General 42, 285309 (2009).

\bibitem{AMRS}
G. Auberson , G. Mahoux , S.M. Roy, Virendra Singh,  
arXiv:quant-ph/0205157 and Phys Lett A300,p.327-333(2002);
  {\it Journ. Math. Phys.} \underbar{44}, 2729-2747 (2003), and \underbar{45},4832-4854 (2004).
  
\bibitem{Grangier}
J. Wenger, M. Hafezi, F. Grosshans, R. Tualle-Brouri and P. Grangier. ''Max-
imal Violation of Bell Inequalities using Continuous Variables Measurements``,
arXiv:quant-ph/0211067 ,and Phys. Rev. A {\bf 67}, 012105 (2003). 
  
  \bibitem{modular}
 Y. Aharonov, H. Pendleton, and A. Petersen, Int. J.
Theor. Phys. {\bf 2}, 213 (1969);S. Massar and S. Pironio, Phys. Rev. A {\bf 64}, 062108 (2001);
  Y. Aharonov, D. Rohrlich, Quantum Paradoxes, Wiley-VCH, (2005) ;S. Popescu, Nat. Phys. {\bf 6}, 151 (2010).

  \bibitem{Cabello}
  A. R. Plastino and A. Cabello, " State-independent quantum contextuality for continuous variables",
  Phys. Rev. A {\bf 82}, 022114 (2010), and arXiv:1005.1620 [quant-ph]; A. Cabello,  `experimentally 
testable state-independent quantum contextuality', Phys. Rev. Lett. 101, 210401 (2008) and arXiv:0808.2456;
   G. Kirchmair, F. Z ̈ahringer, R. Gerritsma, M. Klein-
mann, O. G ̈uhne, A. Cabello, R. Blatt, and C.F. Roos,
Nature (London) 460, 494 (2009); H. Bartosik, J. Klepp, C. Schmitzer, S. Sponar, A. Cabello, H. Rauch, and Y. Hasegawa, 
Phys. Rev. Lett. 103, 040403 (2009); E. Amselem, M. R ̊admark, M. Bourennane, and 
A. Cabello, Phys. Rev. Lett. 103, 160405 (2009); B. H. Liu, Y. F. Huang, Y. X. Gong, F. W. Sun, Y. S.
Zhang, C. F. Li, and G. C. Guo, Phys. Rev. A 80, 044101 (2009); O. Moussa, C. A. Ryan, D. G. Cory, and R. Laflamme,
Phys. Rev. Lett. 104, 160501 (2010); 
A Cabello, M. Kleinmann, C. Budroni,"Necessary and sufficient condition for quantum state-independent contextuality ",
 Phys. Rev. Lett. {\bf 114}, 250402 (2015), and arXiv:1501.03432 [quant-ph].
 
 \bibitem{Walborn}
 C. Gneiting and K. Hornberger, Phys. Rev. Lett.
106, 210501 (2011); M. A. D. Carvalho, J. Ferraz, G. F. Borges, P.-L. de Assis,
S. Pádua, and S. P. Walborn, Phys. Rev. A 86, 032332 (2012); M. R. Barros, O. J. Farías, A. Keller, T. Coudreau,
P. Milman, and S. P. Walborn, Phys. Rev. A 92, 022308 (2015).

  
\bibitem{Asadian}
A. Asadian, C. Budroni, F.E. Steinhoff, P. Rabl, and O. Gühne, O. , "Contextuality in phase space",
Physical review letters, 114(25), 250403 (2015) and  arXiv:1502.05799 [quant-ph];
A. Asadian, C. Brukner, P. Rabl , "Probing macroscopic realism via Ramsey correlations measurements",
Phys. Rev. Lett. 112, 190402 (2014), and arXiv:1309.2229 [quant-ph];
Atul S. Arora, Ali Asadian, "Proposal for a macroscopic test of local realism with phase-space measurements",
 PhysRev A. {\bf92}.062107 (2015).
    
\bibitem{Ketterer}
A. Ketterer, S. P. Walborn, A. Keller, T. Coudreau, and P. Milman, arXiv:1406.6388;
P. Vernaz-Gris, A. Ketterer, A. Keller, S. P. Walborn,
T. Coudreau, and P. Milman, Phys. Rev. A 89, 052311 (2014);
Andreas Ketterer, "Modular variables in quantum information", 
Ph. D. Thesis (October 2016),https://www.researchgate.net/publication/316828699 .
    

\bibitem{Bell1986}
J. S. Bell, Ann. (N.Y.) Acad. Sci. 480, 263 (1986).



\bibitem{Cirel'son}
B. S. Cirel'son,Lett. Math. Phys. {bf 4},93 (1980); B. S. Tsirel'son,
J. Sov. Math. {bf 36},557 (1987).


\bibitem{Werner}
R. F. Werner and M. M. Wolf,Phys. Rev. {\bf 64 A},032112(2001).




\bibitem{Gradshteyn}
I. S. Gradshteyn and I. M. Ryzhik, 'tables of integrals, series and products',Eq 0.234(2), p.7, (Academic Press, San Diego,1980).

  


\end{thebibliography}
\end{document}